\documentstyle[epsf,12pt,a4]{article}
\newcommand{\be}{\begin{equation}}
\newcommand{\ba}{\begin{eqnarray}}
\newcommand{\ee}{\end{equation}}
\newcommand{\ea}{\end{eqnarray}}

\newcommand{\GeV}{\;\mbox{GeV}}
\newcommand{\eV}{\;\mbox{eV}}
\newcommand{\cm}{\;\mbox{cm}}

\newcommand{\secs}{\;\mbox{s}}

\newcommand{\simgt}{\stackrel{>}{{}_\sim}}
\newcommand{\ol}{\overline}

\newcounter{currequation}
\begin{document}
\title{
Dark-matter particles and baryons from inflation and spontaneous CP violation
in the early universe
}
\author{Saul Barshay\footnote{barshay@kreyerhoff.de} and Georg Kreyerhoff\footnote{georg@kreyerhoff.de}\\
III.~Physikalisches Institut A\\
RWTH Aachen\\
D-52056 Aachen}
\maketitle
\begin{abstract}
We present aspects of a model which attempts to unify the creation of cold dark matter, 
a CP-violating baryon asymmetry, and also a small, residual vacuum energy density, in
the early universe. The model contains a primary scalar (inflaton) field and a primary
pseudoscalar field, which are initially related by a cosmological, chiral symmetry. The
nonzero vacuum expectation value of the pseudoscalar field spontaneously breaks CP invariance.
\end{abstract}
There is a relevant open question in relation to a hypothetical, brief inflationary period in the
expansion of the early universe, initiated by the energy density of a scalar field. The question
is whether the energy density in cold dark matter, and a very small energy density
in an effective cosmological constant can arise in the early universe
from the fields involved in the dynamics of inflation. Also, one can ask whether a spontaneous CP
violation is related to the vacuum expectation value of a pseudoscalar field involved in the dynamics
of inflation, and whether this dynamics might contain a reason for the actual, small value of
the baryon to photon ratio, a few times $10^{-10}$ (alternatively, for the empirical number
of baryons, a few times $10^{78}$). In this paper, we describe several properties of a model
which contains a scalar inflaton field with a large vacuum expectation value \cite{ref1,ref2}
at a calculated \cite{ref3,ref1}, potential minimum, and also a related pseudoscalar
field, whose small vacuum expectation value breaks CP invariance spontaneously \cite{ref2}. These properties,
taken together, suggest specific affirmative answers to the above questions.

In contrast to most models for the appearance of radiation and matter near the end of inflation
\cite{ref4,ref5,ref6}, the present model separates into two distinct dynamical mechanisms the
hypothetical origin of primary cold dark matter and the origin of the primary radiation.\footnote{This model
\cite{ref1,ref2} considers dark-matter particles to be not solely a somewhat massive, stable remnant
of the initial radiation, whose present energy density has evolved, rather accidentally, to be about
five times greater than the energy density in ordinary baryons. The model attempts to explain the latter
fact in a unified way, including CP noninvariance leading to the the absolute number of baryons.}
The required ``initial'' energy densities are vastly different: at $\sim 10^{-36}\secs$,
$\sim 10^{36}\GeV^4$ in primary cold dark matter, and $\sim 10^{60}\GeV^4$ in (hot) radiation.
The two are decoupled in this model. Inflation occurs when the classical scalar field $\phi$
is at a calculated \cite{ref3,ref1} potential maximum,\footnote{The illustrative, renormalization-group
calculations which explicitly give rise to an inflaton effective potential with a maximum and a minimum
\cite{ref3,ref1}, involve radiative corrections to a $\lambda\phi^4$ potential in a model with 
self-coupled scalar $(\phi)$ and pseudoscalar $(b)$ fields, and also a massive lepton (which decays into
radiation) with a scalar coupling to $\phi$ and a pseudoscalar coupling to $b$.} at \cite{ref1} the
Planck energy scale $M_P=1.2\times 10^{19}\GeV$, or just above \cite{ref3} this scale. Inflation approaches termination as $\phi$ ``falls''
to a calculated \cite{ref3,ref1} potential minimum,$^{F2}$ taken here at about $\phi=\phi_c
=10^{18}\GeV$. Consider, as an example of the mechanism, that bosonic quanta $(S)$ with mass
$m_S \cong 10^{15}\GeV$, which subsequently decay into radiation with an energy density of the
order of $m_S^4 \cong 10^{60}\GeV^4$, can be produced by a coupling to the energy of motion
$\dot{\phi}^2/2$ of the inflaton field, as its potential energy density is diminishing. It
is noteworthy that a typical, initial \cite{ref1,ref3} potential energy density is of this
order of magnitude; in particular, $\lambda\phi^4 \to \sim 6\times 10^{62} \GeV^4$ for 
$\phi\to \sim M_P$ and with  a small self-coupling parameter $\lambda \sim 3\times 10^{-14}$
\cite{ref7,ref6}. An estimated order of magnitude of the time for conversion of the initial,
inflaton potential (``vacuum'') energy into $S$ quanta is $\sim (1/m_S)(\phi_c/m_S)
\cong 0.7\times 10^{-36}\secs$. A time for $S$ decay has been estimated \cite{ref2} to be $\sim 10^{-36}\secs$
(for $m_S \cong 10^{15}\GeV$); this is then the approximate time for the appearance of 
radiation.\footnote{The uniform, effective radiation temperature is then fixed by $m_S$ to be $\sim 10^{15}\GeV$.
The Hubble expansion rate is of the order of the decay rate for $S\to {\mathrm{radiation}}$. Therefore,
these decay processes do not occur in equilibrium conditions.} 
The entropy in the volume $V\sim (4\pi/3) (3\cm)^3$, which evolves to the presently
observable universe (with radial dimension taken here as $r_0 \sim 1.5\times 10^{28} \cm$) is then
about $1.4 \times 10^{88}$. The simplest estimate for evolution to the present epoch gives the radiation
energy density of approximately $(10^{60}\GeV^4) (10^{-36}/10^{12})^2 (10^{12}/4\times 10^{17})^{8/3}
\sim 1.25\times 10^{-51}\GeV^4$, when $\sim 10^{12}\secs$ is used as the approximate time for matter
dominance, and $\sim 4\times 10^{17}\secs$ as the present age.\footnote{Empirically, including (not yet
observed) neutrinos, it is $\sim 3\times 10^{-51}\GeV^4$. } 
The
above sequence for the appearance of radiation is meant to be illustrative of a possibility.
Our main concerns in this paper
are the creation of cold dark matter and the origin of the baryon asymmetry, to which we turn. 

Consider the production of cold dark-matter particles near to the end of inflation, just before 
$10^{-36}\secs$ (just prior to $S$ decay, the appearance of radiation). Without inflation, the causal horizon
would be at a dimension of $\sim 3\times 10^{-26}\cm$; thus there would be $\sim 10^{78}$ causally
disconnected volume elements within $V\sim (4\pi/3) (3\cm)^3 \cong 113\cm^3$. Inflation of scale by
$\sim 10^{26} (\sim e^{60})$ rapidly expands the causal dimension to cover the dimension of
about $3\cm$, which corresponds to the radial size of the the region at $10^{-36}\secs$ which subsequently
expands to the presently observable universe. Near to the end of inflation, dark-matter particles $\phi$
(inflatons) with an estimated \cite{ref2,ref1} mass $m_\phi \sim 5\times 10^{11}\GeV (=2\sqrt{2}\sqrt{\lambda}
\phi_c)$, can be created by a time-varying gravitational field.\cite{ref5,ref6} This mass is determined
by $\phi_c$ and $\lambda$, independently of the possible contribution to the dark-matter energy
density, and independently of the Hubble expansion rate. The Hubble expansion rate is $H\sim 1/10^{-36}\secs
= 6.6 \times 10^{11}\GeV$. Thus $H$ is $\simgt m_\phi$.
Stated alternatively, $VH^4/m_\phi\sim 10^{78}$. Gravitational production of essentially
a condensate of $\phi$ particles can produce a number density $n_\phi \sim \epsilon H^3$, and an
energy density $\rho_\phi \sim \epsilon H^3 m_\phi$, where $\epsilon \cong (1/12\pi^2)
(1-6\xi)^2$.\cite{ref5} The factor $(1-6\xi)$ is related to a hypothetical small \cite{ref5} deviation
from conformal invariance for field and (massless) particle coupling. For creation of $\phi$ particles,
we consider this factor to be representative of an effective suppression factor, when $m_\phi\sim H$.
Taking as
an example of an effective suppression factor, $|1-6\xi|\sim 1.7\times 10^{-5}$, gives 
$\epsilon \sim 7.5\times 10^{-12}$, a number similar to that previously estimated \cite{ref1}
for small field fluctuations; one then obtains $\sim 3\times 10^{67}$ dark-matter $\phi$
particles. 
Their
(metastable \cite{ref1}) presence would  give a present cold dark-matter energy density of
about $1.5\times 10^{79}\GeV/ (4\pi/3) (1.5\times 10^{28}\cm)^3 \cong 0.9 \times 10^{-47} \GeV^4$,
which is about $23\%$ of critical, in agreement with cosmological data.\cite{ref8}\footnote{There is
a second possible mechanism for production of $\phi$ particles, which involves the self-coupling of the
$\phi$ field and small fluctuations about $\phi=\phi_c$, as is illustrated in reference 1. If most
of the energy has already gone into the creation of $S$ particles, these later fluctuations will be suppressed. Note
that if $(1-6\xi)$ fluctuates around zero, then terms of \cite{ref5} of order of $\pm |1-6\xi|^3$ could
represent small primordial density fluctuations around a mean density proportional to $(1-6\xi)^2$.}

The cosmological, chiral model \cite{ref2} contains a pseudoscalar field $b$, in addition to the
scalar inflaton field $\phi$. When the chiral symmetry is spontaneously broken at $\phi=\phi_c$, the quanta
of the $b$ field are the necessary massless Goldstone bosons,\cite{ref9} just as the pions are
the Goldstone bosons in the $\sigma$-model for nucleon (equivalently, constituent quark) mass in low-energy
particle physics.\cite{ref10,ref11} We have examined some consequences \cite{ref2, ref12} of the hypothesis
that the vacuum expectation value of the $b$ field has a nonzero value $F_b$. CP invariance is then
spontaneously broken.\cite{ref13} The small scale of $F_b\sim 5.5\eV$ is estimated independently
by the hypothesis \cite{ref2} that it gives rise to mass for a neutrino.\footnote{
In particular, the mass of the heaviest ordinary neutrino, presumably $\nu_\tau$, for which
experiments suggest $m_{\nu_\tau} \sim 0.055\eV$ (assuming no near-degeneracy of neutrino masses).}
Two results \cite{ref2} following from $F_b\neq 0$ provide motivation for the hypothesis.
\begin{itemize}
\item[(1)] A significant asymmetry between the number of antineutrinos and the number of neutrinos
can be generated in the early universe in a non-equilibrium, primary decay process, involving
necessary, CP-violating final-state interactions from exchange of $b$ quanta. 
This asymmetry might be partly transformed into a baryon asymmetry at the time of electroweak
symmetry breaking.
Below, we try to explain the empirical numerical value of the baryon asymmetry, from its
possible seeding by a primary antineutrino-neutrino asymmetry.
\item[(2)] There is a small vacuum energy density associated with the $b$ field, of magnitude
$|-\lambda F_b^4| \cong 2.7 \times 10^{-47}\GeV^4$.\footnote{A possible reason for this residual vacuum
energy (effective cosmological constant) being empirically a positive number, is given in the appendix
of reference 12. In line 9, for the misprinted $\sqrt{\lambda}$ read $\lambda$.} This is explicitly related to a small
neutrino mass, through $F_b$. Clearly, the very small value of $\lambda$ for the self-coupling of the
bosons in the chiral model, also is crucial for the size of this effective cosmological constant.
\end{itemize}

In the evolution from $\sim 10^{-36}\secs$ to the time of electroweak symmetry breaking, $\sim 10^{-12}\secs$,
the radial dimension of $\sim 3\cm$ expands to $3\times 10^{12}\cm$. An intrinsic baryon dimension is 
$1/m_{\mathrm{baryon}}\cong 2\times 10^{-14}\cm$. There are $\sim 3\times 10^{78}$ such volume elements.
With $\sim 3\times 10^{87}$ neutrinos (of one flavor) in the radiation, a CP-violating asymmetry in number
density produced in an early $S$ decay mode of, for example $(n_{\ol{\nu}}-n_\nu)/n_\nu \sim 10^{-9}$,
as has been estimated \cite{ref2}, results in $3\times 10^{78}$ excess $\ol{\nu}$, or about one per volume
element with $r_{\mathrm{baryon}} \sim 1/m_{\mathrm{baryon}}$. The hypothetical \cite{ref14,ref15}
transitions $\ol{\nu}\to 3$ quarks occur in each volume element; this could result in $\sim 3\times 10^{78}$
baryons when the confinement transition occurs at $\sim 10^{-6}\secs$, where baryons form from three
quarks.\footnote{It is assumed that the spatial domain for three energetic, correlated quarks which are
subsequently to be confined by QCD as valence quarks in a single baryon is also characterized by $r_{\mathrm{baryon}}$. 
Note that a dimension  $r_{\mathrm{baryon}}\sim 2\times 10^{-14}\cm$ corresponds to expansion of the
dimension $1/m_\phi$ by the factor $10^{12}$ in going from $\sim 10^{-36}\secs$ to $\sim 10^{-12}\secs$.
} At this time, the energy densities of baryons and dark matter, $\rho_{\mathrm{baryon}}$
and $\rho_{\mathrm{dm}}$ are fixed. We have the ratio $\rho_{\mathrm{baryon}}/\rho_{\mathrm{dm}}
\sim 3\times 10^{78}\GeV/1.5\times 10^{79}\GeV \sim 20\%$; thereafter both energy densities fall like
matter densities to the present. The model thus contains a tentative explanation for the rather
small absolute number of baryons. The ratio of baryon to dark-matter energy densities agrees well with
cosmological and other data \cite{ref8}. Of course, the primary antineutrino-neutrino asymmetry could be
larger than the $\sim 10^{-9}$ used in the above numerical example.\cite{ref2} The suggestion
here is that then, very many more than $\sim 3\times 10^{78}$ baryons cannot be accommodated;
many more will not form within the available $\sim 3\times 10^{78}$ volume elements with
$r_{\mathrm{baryon}} \sim 1/m_{\mathrm{baryon}}$. (There are arguments that the baryon asymmetry
could have been much larger, say up to one hundred times, with reference to ``a priori'' permissable
amounts of primordial ${\mathrm{He}}^4$ and ${\mathrm{H}}^1$.\cite{ref16}) In the present model,
there is a natural expectation of a residual antineutrino-neutrino asymmetry; this asymmetry
could well be larger than the baryon asymmetry.\cite{ref2}

Before concluding, we discuss consequences of a speculation that the (metastable) dark-matter
particles are not the very massive, scalar inflaton quanta $\phi$,\cite{ref1} but rather the
Goldstone boson-like pseudoscalar quanta $b$. When $F_b\neq 0$, the $b$ particles have a quite
small mass, estimated \cite{ref2} to be $m_b = 2 \sqrt{2}\sqrt{\lambda} F_b \sim
2.7 \times 10^{-6}\eV$. The chiral-like \footnote{By chiral-like, we mean that although the symmetry 
is also broken explicitly (in particular, note footnote b in reference 12),  
important properties of the spontaneous breaking persist: the related existence of a very massive, scalar
particle and a very light, pseudoscalar particle.} model for related (explicitly through $\lambda$)
scalar and pseudoscalar quanta in the early universe thus contains both a high energy scale $\phi_c$
(or $m_\phi$) and a small energy scale $F_b$ (or $m_b$), with the large ratio
$(m_\phi/m_b) = (\phi_c/F_b) \cong 1.85 \times 10^{26}$. Consider the hypothesis that the
condensate of $\phi$ particles becomes effectively a condensate of $b$ particles
with of order $(\phi_c/F_b)$ $b$ particles per $\phi$ particle. The primary energy density in dark-matter
particles is unchanged. The total number of dark-matter degrees of freedom increases, from
$\sim 3\times 10^{67}$ to $\sim 5.5\times 10^{93}$.
This could further ensure dark-matter smoothness.\cite{ref12}\footnote{There is an attractive interaction between a pair
of very-slowly moving $b$ particles, via $b$ exchange, which is characterized by $1/m_b \sim 7\cm$,
as calculated
in reference 12. An intrinisic dimension for a $b$ particle of the order of $\lambda^2/m_b \sim 0.7\times
10^{-26}\cm$ is close to $1/m_\phi$. Thus $\lambda^2(\phi_c/F_b)\sim 1$ relates the three, a priori
independent, dynamical variables.} Only small fluctuations in the radiation temperature are induced (see
Appendix).
This suggests that 
fluctuations may become smaller at the largest angular scales (the largest dimensions). This might be
a feature of the temperature fluctuation measurements in the CMB.\cite{ref8}

In summary, we have discussed certain positive aspects of a model which attempts to unify the creation
of cold dark matter, a CP-violating baryon asymmetry, and also a small residual vacuum energy
density, in the early universe. Predictions include a significant antineutrino-neutrino asymmetry,
and the possibility of a diminution of CMB temperature fluctuations at the largest angular scales.
As in all such models,\cite{ref4} it is difficult to test experimentally in an unambiguous way,
particle-physics pictures of the earliest time interval in the universe. One tends to be somewhat
reliant upon the possibly circumstantial representation of important global facts, as in the
present model. Despite this reservation, two essential ideas which the model incorporates,
seem worthwhile to explore. One idea is that spontaneous CP violation arises from the nonzero
vacuum expectation value of a primary, Goldstone boson-like pseudoscalar field related to the
inflaton field, with the smallness of the vacuum expectation value being related to a small
neutrino mass. A second idea is that of different dynamical mechanisms for the creation of cold
dark-matter particles, and for the creation of radiation. If $b$ particles constitute cold
dark matter, then both dark-matter mass and an effective cosmological constant arise from
the same field variable $F_b$, which variable gives spontaneous CP violation leading to the baryon
asymmetry. This is a dynamical unification.

\section*{Appendix}

The empirical number that characterizes CMB temperature fluctuations on small angular
scales is of the order of $10^{-5}$ at the decoupling time.\cite{ref8} It could have been
significantly larger, apriori. When interpreted as a consequence of fluctuations in the
gravitational potential at the decoupling time, it is perhaps interesting that the
presence then of $N$ nearly-distinct aggregates of dark matter with mass of order
$5\times 10^{69}\GeV$ and with dimension of about $10^{22}\cm$, within a region with
dimension of about $10^{24}\cm$, can give a (Gaussian) temperature fluctuation \cite{ref4}
of the order of $1/30 (1/\sqrt{N}) \sim 1/30 (10^{-3}) \sim 3\times 10^{-5}$, for a maximum
$N$ determined by available volume, $N\sim (10^{24}/10^{22})^3 = 10^6$. It is noteworthy
that such a hypothetical dark-matter ``pre-galaxy'' can have a mean energy density which
becomes greater than a falling ``background'' dark-matter energy density, somewhat after the
decoupling time. Visible galaxies would tend to form early from stars in these seed
systems (in about $10^8$ yr). Rich $(\sim \sqrt{N}\sim 10^3)$ galactic clusters with in-falling,
KeV-temperature cluster gas, form later (in $>10^9$ yr), as observations indicate.\cite{ref17} Also,
there is a suggestion here that mass correlations even greater than of order $10^{15}$ solar masses
(superclusters) can develop from the early seeds.

Finally we note that from the relation in the last sentence of footnote F10, the present ideas could
possibly contain a reason for the usual postulated \cite{ref4}, empirically necessary, small value
pf the dimensionless parameter $\lambda$. This is that $\lambda^2$ is given by the small ratio
of the two vacuum expectation values, for Goldstone-like pseudoscalar field and inflaton
scalar field, which fields constitute the basis of the inflation model with a spontaneously-broken
CP invariance: $\lambda \sim \sqrt{F_b/\phi_c} \sim 10^{-13}$, with $F_b$ related to a residual
vacuum energy density and to neutrino mass.

We thank the referee for helpful questions.


\begin{thebibliography}{9}
\bibitem{ref1} S.~Barshay and G.~Kreyerhoff, Eur.~Phys.~J.~{\bf C5},  369 (1998)
\bibitem{ref2} S.~Barshay and G.~Kreyerhoff, Mod.~Phys.~Lett. {\bf A19}  2899  (2004)
\bibitem{ref3} S.~Barshay and G.~Kreyerhoff, Z.~Phys.~{\bf C75},  167 (1997); Erratum: ibid.~{\bf C76}, 577 (1997)
\bibitem{ref4} E.~W.~Kolb and M.~S.~Turner, {\it The Early Universe}, Addison-Wesley Publishing Co.~1990, chapter 8
for inflation and original references; chapter 9.6.2 for CMB fluctuations.
\bibitem{ref5} L.~H.~Ford, Phys.~Rev.~{\bf D35}  2955 (1986)
\bibitem{ref6} P.~J.~E.~Peebles and A.~Vilenkin, Phys.~Rev.~{\bf D59} 063505 (1999)
\bibitem{ref7} H.~Kurki-Suonio and G.~J.~Mathews, Phys.~Rev.~{\bf D50}  5431 (1994)
\bibitem{ref8} WMAP Collaboration, C.~L.~Bennett et al., astro-ph/0302207v3, and references therein
\bibitem{ref9} J.~Goldstone, Nuovo Cimento {\bf 19} 155 (1961)
\bibitem{ref10} M.~Gell-Mann and M.~Levy, Nuovo Cimento {\bf 16} 705 (1960)
\bibitem{ref11}  B.~W.~Lee, {\it Chiral Dynamics}, Gordon and Breach Science Publishers, Inc., 1972
\bibitem{ref12} S.~Barshay and G.~Kreyerhoff, Mod.~Phys.~Lett.~{\bf A20}  1155 (2005)
\bibitem{ref13} T.~D.~Lee, Phys.~Rep.~{\bf 9C}  143 (1974);\\
	T.~D.~Lee, {\it Particle Physics and Introduction to Field Theory}, Harwood Academic Publishers,
1981, chapter 16
\bibitem{ref14} G.~'t Hooft, Phys.~Rev.~Lett.~{\bf 37}  8 (1976);\\
	Phys.~Rev.~{\bf D14} 3432 (1976)
\bibitem{ref15} V.~A.~Kuzmin, V.~A.~Rubakov and M.~E.~Shaposhnikov, Phys.~Lett.~{\bf 155} 36 (1985)\\
	V.~A.~Rubakov, Nucl.~Phys.~{\bf B256} 509 (1985)
\bibitem{ref16} D.~V.~Nanopoulos, TH.~2738-CERN, 1979
\bibitem{ref17} J.~P.~~Henry, U.~G.~Briel and H.~B\"ohringer, Scientific American, Dec.~1998, 24
\end{thebibliography}
\end{document}